# Respective influence of in-plane and out-of-plane spin-transfer torques in magnetization switching of perpendicular magnetic tunnel junctions

A.A. Timopheev[1,2,3], R.Sousa[1,2,3], M.Chshiev[1,2,3], L.D. Buda-Prejbeanu[1,2,3],

B. Dieny[1,2,3]

1. Univ. Grenoble Alpes, INAC-SPINTEC, F-38000 Grenoble, France
2. CEA, INAC-SPINTEC, F-38000 Grenoble, France
3. CNRS, SPINTEC, F-38000 Grenoble, France

Abstract

The relative contributions of in-plane (damping-like) and out-of-plane (field-like) spin-transfer-torques in the magnetization switching of out-of-plane magnetized magnetic tunnel junctions (pMTJ) has been theoretically analyzed using the transformed Landau-Lifshitz (LL) equation with the STT terms. It is demonstrated that in a pMTJ structure obeying macrospin dynamics, the out-of-plane torque influences the precession frequency but it does not contribute significantly to the STT switching process (in particular to the switching time and switching current density), which is mostly determined by the in-plane STT contribution. This conclusion is confirmed by finite temperature and finite writing pulse macrospin simulations of the current-field switching diagrams. It contrasts with the case of STT-switching in in-plane magnetized MTJ in which the field-like term also influences the switching critical current. This theoretical analysis was successfully applied to the interpretation of voltage-field STT switching diagrams experimentally measured on perpendicular MTJ pillars 36 nm in diameter, which exhibit macrospin-like behavior. The physical nonequivalence of Landau and Gilbert dissipation terms in presence of STT-induced dynamics is also discussed.

1. Introduction

Fully perpendicular magnetic tunnel junctions (pMTJ) constitute the storage element of spin-transfer-torque magnetoresistive random access memory (STT-MRAM) [1-6]. STT-MRAM are very promising emerging non-volatile memories since they combine non-volatility, low energy consumption, high thermal stability and almost unlimited endurance. The strongest research and development efforts are nowadays focused on out-of-plane magnetized MgO-based MTJs. Indeed, the latter combine several advantages.They exhibit a high tunnel magnetoresistance effect [7] amplitude due to a very efficient spin-filtering phenomenon associated with the symmetry of the tunneling electron wave function [8,9]. Furthermore, they present a very large interfacial perpendicular anisotropy at the interface between the magnetic electrode and the MgO oxide barrier (Ks~1.4erg/cm²) [10] which allows to achieve a quite high thermal stability of the storage layer magnetization and therefore a long memory retention. In addition, a remarkable property of this interfacial anisotropy is that it exists in materials having weak spin-orbit coupling and therefore relatively low Gilbert damping $\alpha$ ($\alpha$<0.01). This is very important in STT-MRAM since the critical current for STT-induced switching [11,12] of the storage layer magnetization is directly proportional to the Gilbert damping. The advantage of using out-of-plane magnetized MTJs in STT-MRAM rather than in plane ones is twofold: firstly, the interfacial perpendicular anisotropy at CoFeB/MgO interface provides higher thermal stability at smaller dimensions (sub-60nm) than the usual shape anisotropy provided by giving elliptical shape to in-plane magnetized MTJs. Secondly, for a given retention i.e. a given thermal stability factor, the critical current for STT-induced switching is lower with out-of-plane magnetized storage layer than it is for an in-plane magnetized one [13,14].

From a theoretical point of view, a first approach to STT-induced switching can be developed by solving the Landau Lifshitz Gilbert equation under the assumptions of zero kelvin macrospin approximation under stationary applied spin-polarized current. The equilibrium configurations of the system can thus be calculated and the precessional dynamics of the system submitted to a small perturbation from the static equilibrium can be studied. This allows to derive the threshold current required to achieve STT switching, as it was done in Refs. [13-15]. Thermal fluctuations can be taken into account in several limiting cases using Fokker-Planck equation. Thermal activation mainly decreases the threshold current value and the switching time introducing an undesirable effect of stochasticity in magnitude of both parameters [16,17]. The influence of the writing pulse duration was also theoretically studied [16,18-21].

Despite the numerous experimental results [22, 23] and micromagnetic simulations [24-26] generally pointing on quantitative disagreements with the macrospin-based estimations, usage of the macrospin approach is still justified for at least for two reasons. First of all, it gives a simple but solid picture of the physical processes involved in the STT switching that creates a common basis for qualitative analysis of the different magnetic multilayered systems, while most of the conclusions derived from micromagnetic approaches are rather of particular character. Micromagnetic behavior can be mimicked, for example, by introduction of an effective activation volume instead of Stoner-Wohlfarth behavior, but still using a thermal activation model for the subvolume [22]. Secondly, considering the general trend to reduce the volume of the storage element (and, consequently, the energy needed per write/read cycle), magnetic memory elements will eventually behave in a macrospin manner.

Based on these viewpoints, we investigated the STT switching in fully perpendicular magnetic tunnel junction systems, where in addition to Slonczewski STT term (sometime called in-plane torque since it lies in the plane defined by the local magnetization and that of the spin-polarization usually defined by the magnetization direction of the reference pinned layer), having damping-like structure, an out-of-plane, or field-like term exists. Several theoretical works predicted that the torque produced by out-of-plane STT term could reach an amplitude comparable to that of in-plane torque [27-29]. Several experimental works carried out on in-plane MTJ structures have already estimated it to be in the range of 30-40 % of the in-plane torque [30-33]. It was mentioned [34] that its presence may lead to a backswitching process, a very undesirable effect in magnetic memory applications causing write errors.

In this study, after having analyzed the Landau-Lifshitz-Gilbert-Slonczewski equation mathematically transformed into Landau-Lifshitz form, we show that in fully perpendicular MTJ structures, the field-like torque plays a negligible role in the switching process. In contrast to in-plane MTJ systems [30-34], it only influences the precessional frequency preceding the switching but the switching current density is primarily determined by the in-plane STT term. The experiment carried out on 36nm diameter pMTJ pillar supports our conclusions.

## 2. Phase boundaries from LLG equation transformed into LL equation

The most accepted form of LLG equation describing dynamics of a macrospin under constant spin polarized current can be presented as follows:

$$\frac{d\hat{m}}{dt} = -\gamma(\hat{m} \times \vec{H}_{eff}) + \alpha\left(\hat{m} \times \frac{d\hat{m}}{dt}\right) - \gamma\,\hat{m} \times (\hat{m} \times a_\parallel\,\hat{p}) + \gamma\,\hat{m} \times a_\perp\,\hat{p}\,, \quad (1)$$

here $\hat{m} = \frac{\vec{M}}{M_S}$ – unit vector along the free layer magnetization direction ($M_S$ – free layer's volume magnetization saturation parameter), $\vec{H}_{eff}$ effective field (comprising applied field, anisotropy field, demagnetizing field), $\hat{p}$ – unit vector along the polarizer layer magnetization direction, $\alpha$ Gilbert damping, $\gamma$ gyromagnetic ratio. $a_\parallel$ and $a_\perp$ are, respectively, in-plane (damping-like) and out-of-plane (field-like) spin-transfer-torque prefactors. Both prefactors can be phenomenologically represented as functions of the spin polarization in the magnetic electrodes, current density or voltage bias applied to the tunneling barrier as will be done later in the text.

In-plane and out-of-plane STT terms as written in Eq. (1) are geometrically equivalent to the precession and damping terms of Landau-Lifshitz equation. One can therefore transform Eq. (1) into Landau-Lifshitz form using the standard technique, i.e. by making a $\hat{m} \times$ product on both sides of equation:

$$\widehat{m} \times \frac{d\widehat{m}}{dt} = -\gamma\, \widehat{m} \times (\widehat{m} \times \vec{H}_{eff}) + \alpha\, \widehat{m} \times \left(\widehat{m} \times \frac{d\widehat{m}}{dt}\right) - \gamma\, a_\parallel\, \widehat{m} \times (\widehat{m} \times (\widehat{m} \times \widehat{p})) + \gamma\, a_\perp \widehat{m} \times (\widehat{m} \times \widehat{p}),$$

and putting obtained in a replacement of the damping term in Eq.(1). This yields,

$$\frac{(1+\alpha^2)}{\gamma} \frac{d\widehat{m}}{dt} = -\widehat{m} \times \left(\vec{H}_{eff} - (a_\perp + \alpha\, a_\parallel)\, \widehat{p}\right) - \widehat{m} \times \left(\widehat{m} \times \left(\alpha\, \vec{H}_{eff} - (\alpha\, a_\perp - a_\parallel)\, \widehat{p}\right)\right). \quad (2)$$

To this moment, all the transformations born only a character of mathematical identities and Eq. 2 is valid for any system with any configuration of $\vec{H}_{eff}$ and $\widehat{p}$. Rewritten in such a way, it acquires more suitable form for further analytical treatment because dynamics in this system is fully determined by two vectors, namely $\left(\vec{H}_{eff} - (a_\perp + \alpha\, a_\parallel)\, \widehat{p}\right)$ and $\left(\alpha\, \vec{H}_{eff} - (\alpha\, a_\perp - a_\parallel)\, \widehat{p}\right)$, which have many similarities and whose form can be significantly simplified as soon as actual geometry of $\vec{H}_{eff}$ and $\widehat{p}$ has been set. Also Eq. (2) is more convenient to use in numerical integration schemes. Further analysis will be focused on the case of pMTJ structure assuming macrospin dynamics of the storage layer described by Eq.(2).

We consider fully perpendicular magnetic tunnel junctions submitted to an out-of-plane external magnetic field $\vec{H}_{ext}$ therefore applied parallel to the symmetry axis. This situation allows analytical analysis wherein the quantities $\vec{H}_{eff}$, $\widehat{p}$, $\vec{H}_{ext}$, $\hat{z}$ remain collinear independently of the instantaneous direction of $\widehat{m}$. The magnetic free energy density functional U of such system depends only on one variable $\theta$ – the angle between magnetization vector $\widehat{m}$ and quantization axis $\hat{z}$ (see Fig.1). It writes:

$$U = (K_\perp - 2\pi M_S^2) \sin^2\theta - M_S H_{ext} \cos\theta. \quad (3)$$

When $|H_{ext}| < H_\perp$, $H_\perp = \frac{2K_\perp}{M_S} - 4\pi M_S$, $H_{ext} = \vec{H}_{ext} \cdot \hat{z}$, there are two stable magnetic moment orientations which are independent of $H_{ext}$ and always collinear with $\hat{z}$:

$$\begin{cases} \frac{\partial U}{\partial \theta} = 0,\ \frac{\partial^2 U}{\partial \theta^2} > 0, \\ -H_\perp < H_{ext} < H_\perp, \\ \quad H_\perp > 0. \end{cases} \rightarrow\ \theta_0 = 0,\ \theta_0 = \pi. \quad (4)$$

The collinearity of the four vectors $\vec{H}_{eff}$, $\widehat{p}$, $\vec{H}_{ext}$, $\hat{z}$ yields great simplifications in equation (2), allowing to work only with the magnitudes $a_\perp$, $a_\parallel$ and $H_{eff}$:

$$\begin{aligned} \frac{(1+\alpha^2)}{\gamma} \frac{d\widehat{m}}{dt} &= -\widehat{m} \times A\,\hat{z} - \widehat{m} \times (\widehat{m} \times B\hat{z}), \\ A &= H_{eff} - (a_\perp + \alpha\, a_\parallel), \\ B &= \alpha\, H_{eff} - (\alpha\, a_\perp - a_\parallel), \\ H_{eff} = \vec{H}_{eff} \cdot \hat{z} &= -\frac{\partial U}{\partial \vec{M} \cdot \hat{z}} = H_\perp \left(\cos\theta_0 + H_{ext}/H_\perp\right) \end{aligned} \quad (5)$$

Here, two scalar parameters $A, B$ are introduced, which represent the direction and magnitude of the perpendicular and in-plane (the plane is formed by $\widehat{m}$ and $\widehat{p}$) effective torques (see Fig.1) acting on the magnetization when the latter departs from its equilibrium position $\theta_0$ (0 or $\pi$) because of thermal fluctuations.

Important specifics of the considered system is that $A$-parameter cannot change the orbit (i.e. the angle $\theta$), it only influences the frequency of the precession. One can derive ferromagnetic resonance (FMR) condition, which is just a modified "easy-axis" Kittel's formula for this case:

$$\omega/\gamma = H_{eff} - (a_\perp + \alpha\, a_\parallel), \qquad (6)$$

were $\omega$ is the angular frequency of the resonance precession. One can see, that if $a_\perp > H_{eff} - \alpha\, a_\parallel$ the precession direction will be changed, while increase or decrease of $\theta$ is exclusively determined by the sign of $B$-parameter, wherein the damping-like STT-term is dominating since α is usually small ( typically in the range 0.007 to 0.02). The precessional response of the system before the switching could be measured for instance by measuring $\omega$ versus the DC applied voltage bias, $V_{bias}$, on a single pMTJ pillar either by RF voltage frequency detection, noise measurements [35], spin-torque experiments or by microfocused BLS FMR technique. The excitation frequency would give access to $a_\perp(V_{bias})$ dependence, while the FMR linewidth parameter change versus $V_{bias}$ would reflect mostly the $a_\parallel(V_{bias})$ dependence.

Turning back to the analysis of Eq. (5) and Fig. 1, one can note that only the damping term, $\hat{m} \times (\hat{m} \times B\hat{z})$, can change the precession angle $\theta$. It is therefore possible to derive the boundary conditions for a current-magnetic field stability phase diagram. The magnetization switching process starts when $B$-parameter changes sign. This condition yields the threshold criterion for the STT-induced magnetization switching:

$$\alpha\, H_{eff} + a_\parallel - \alpha\, a_\perp = 0 \qquad (7)$$

One can see from Eq. (7) that the contribution from the in-plane STT term ($a_\parallel$) is largely dominating the switching process. Indeed, the in-plane torque is of the order of $\alpha H_{eff}$ while the contribution of the perpendicular torque is weighted by the Gilbert damping resulting in a much weaker influence in the switching process. Here one can note again that the best method to determine experimentally $a_\perp$ is through FMR measurements, and not from the influence of $a_\perp$ on the (current, field) phase diagram boundaries since the latter is very weak. Indeed, from the above discussion, being able to see an influence of $a_\perp$ on the phase diagram boundaries would require to have $a_\perp \approx a_\parallel/\alpha$ which seems to be physically unachievable in standard pMTJ systems [27-34]. Also, as it will be shown in Sec.6, the $\alpha\, a_\perp$ term in the Eq. (7) disappears if one chooses the dissipation term in the Landau-Lifshitz formulation. In any case, Eqs.(6,7) are quite useful for the analysis of STT switching experiments performed on pMTJ systems.

### 3. Stability phase diagram boundaries

Having set the relations between electric current flowing through pMTJ and the spin-torque prefactors magnitudes, one can construct the stability phase diagram explicitly from Eq. (7) assuming that the spin-polarized current pulse is long enough to complete any STT induced switching while influence of the thermal fluctuations is limited to setting small initial misalignment angle $\theta_0$, so that $|\cos\theta_0| \approx 1$. Modification of the phase boundaries due to thermal fluctuations and under short pulse writing regime, which are essential in real magnetic memory applications, will be analyzed in the following sections, while in this section the conditions of long-pulse and low-temperature regime are assumed.

In most investigated pMTJs, one can expect the condition $a_\perp < a_\parallel$ and $\alpha\, a_\perp \ll a_\parallel$ to be fulfilled. In this case, one can set $a_\perp = 0$ and build up the boundaries of the (current, field) stability phase diagram. In absence of the spin-polarized current ($a_\parallel = 0$, $a_\perp = 0$), the switching occurs when $\alpha\, H_{eff}$ changes sign, i.e. when $H_{ext} = -H_\perp$ for $\theta_0 = 0$ and $H_{ext} = H_\perp$ for $\theta_0 = \pi$. This defines the vertical boundaries on the diagram shown in Fig. 2a, depicted by dashed vertical lines. For $H_{ext} = 0$ and setting $a_\parallel = s_{t\parallel}\, G_p\, V_{bias}$ ($s_{t\parallel} = \frac{\hbar}{2e} \cdot \frac{\eta}{t_F M_s}$ = STT conversion efficiency factor, in units of Oe /(A· cm$^{-2}$); $\eta$ – effective spin polarization parameter; $G_p$ – tunneling conductance factor, generally dependent on $\theta$ and $V_{bias}$, in units of $\Omega^{-1}$cm$^{-2}$, representing in the simplest interpretation the inverse of the RxA product) one can obtain that the switching current density $I_{sw}$ is proportional to $\alpha\, H_\perp$:

$$I_{sw0} = G_p V_{sw0} = \frac{\alpha\, H_\perp}{s_{t\parallel}} = \frac{2e}{\hbar} \cdot \frac{t_F\, \alpha\, H_\perp M_s}{\eta}. \tag{8}$$

In the case $H_{ext} \neq 0$, relation (8) leads to a linear dependence between the switching current and external magnetic field, yielding a linear slope on the switching phase diagram given by:

$$\frac{d\, I_{sw}}{d\, H_{ext}} = \frac{\alpha}{s_{t\parallel}} = \frac{2e}{\hbar} \cdot \frac{t_F\, \alpha\, M_s}{\eta}. \tag{9}$$

One can conclude that if the effective spin polarization parameter $\eta$ is constant (i.e. weakly dependent on the bias voltage $V_{bias}$), then the STT driven parts of the switching diagram is linearly dependent on the applied field with the slope proportional to the intrinsic damping parameter $\alpha$ and inversely proportional to the STT efficiency prefactor $s_{t\parallel}$ and with the zero-field switching current magnitude being proportional to the effective perpendicular anisotropy $H_\perp$. One should also note that Eq. (8) is in full agreement with other previously obtained expressions [13-15, 36] for the zero field threshold switching current derived from the analysis of precessional response of the system, assuming linear dependence of the damping-like STT prefactor versus the applied current. In our case, Eq.(7) allows one to calculate I-H stability phase diagram boundaries for any $a_\parallel$, $a_\perp$ prefactors with arbitrary bias current (voltage) dependence, or by choosing it from the theoretical estimations made for the concrete MTJ system [28,29].

Simultaneous influence of both in-plane and out-of-plane STT terms on the phase boundaries is shown on Fig.2b. We have chosen realistic values for the magnetic system (see the figure caption) letting the in-plane prefactor be linearly dependent on bias voltage with $s_{t\parallel}$ = 67· $G_p^{-1}$ Oe/Volt. As for the out-of-plane prefactor, $a_\perp$, we show three different cases: zero, quadratic dependence with $s_{t\perp 2}$ = 154· $G_p^{-2}$ Oe/(Volt)$^2$ and the third one – quadratic + linear dependence (which mimics features of an asymmetric MTJ structure, see expression in caption of Fig.2) with an unreasonably large STT conversion coefficients. One can see, that within $-H_\perp < H_{ext} < H_\perp$ the difference between the phase boundaries in all three cases is negligible. The second case uses exactly the same parameters as the ones used in Ref.[15] in Fig. 3. We can see that the boundaries calculated and simulated there are identical to all our three cases: no matter what kind of prefactor dependence is introduced for the out-of-plane STT term. This confirms that the out-of-plane STT term has a negligible influence on the STT switching diagram. Parabolic shape of the boundaries starts being observed only in the third case and it becomes noticeably different only for current magnitudes several times larger than the threshold switching current. Thus, one can conclude that under long-pulse/low-temperature conditions, the STT switching in fully perpendicular MTJ structures obeying macrospin dynamics is almost not influenced by the out-of-plane STT term and by its actual prefactor bias voltage or current dependence. Below, we will show that this statement is still valid at finite temperature and reasonably short writing pulses.

### 4. Macrospin simulations

Aiming at extending the conclusions made in the previous sections to the case of finite temperatures and finite writing pulse regime, a series of macrospin simulations were performed using Eq. (2) (i.e. with Gilbert damping). The simulations were carried out with a fixed writing pulse duration of 40 ns and a cumulative integration time of 1 µs for each field point. The following assumptions of bias voltage dependences for the STT prefactors were used: $a_\perp = s_{t\perp 2} G_p^2 V_{bias}^2$ and $a_\parallel = s_{t\parallel} G_p V_{bias}$, which is the case of symmetrical MTJ systems with high spin polarization parameter. For convenience, the parameter $G_p$ was set constant equal to 1 $\Omega^{-1} cm^{-2}$. The temperature was included in the form of stochastic thermal field $H_{th}$ with Gaussian distribution [37], added directly to the effective field $H_{eff}$. Statistical properties of these thermal fluctuations are given by the following relations:

$$\langle H_{th,i}(t) \rangle = 0$$

and

$$\langle H_{th,i}(t) H_{th,j}(t') \rangle = \frac{2\alpha k_B T}{\gamma M_s V_p} \delta_{ij} \delta(t-t')$$

where $k_B$ is the Boltzman constant, and $V_p$ the free layer volume. The chosen LLG equation is integrated with a (predictor-corrector) Heun scheme [38]. Here we used $V_p = 2.07 \times 10^{-17}$ cm$^3$, $H_\perp$ = 200 Oe, $M_s$ = 1000 emu/cm$^3$, which gives the effective stability factor at T = 300 K:

$$\Delta = \frac{H_\perp M_s V_p}{2 k_B T} = 50.$$

This set of the parameters was chosen to mimic working conditions of an actual STT-MRAM device. Two sets of macrospin simulations, at T=0K and T=300K respectively, presented on Fig.3 show how the phase boundaries are changed for the different combinations of in-plane and out-of-plane STT-term prefactors magnitudes. We will discuss firstly the results shown in Fig. 3a corresponding to the case with finite pulse duration and no thermal fluctuations (T=0K).

The finite duration of the writing pulse brings two main effects. Firstly, the STT-driven boundaries are shifted towards much higher voltages (currents). Evidently, to achieve switching within the considered finite time period, one has to apply higher amplitudes for the writing pulses. On the initial stage, when $\hat{m}$ is almost collinear with the symmetry axis $\hat{z}$, the torque is very weak which results in a very slow STT induced dynamics in the system. It is evident that in absence of thermal fluctuations, the switching time from $\hat{m} \parallel \hat{z}$ initial would be infinite for any spin-polarized current magnitude [13,14]. To avoid this in the T=0K simulations, a small misorientation (0.1°) between $\hat{p}$ and $H_{ext}$ was introduced in the system. The second effect is nonlinearities of the phase boundaries which are seen even on the diagrams with the in-plane STT-term only. This effect is linked with non-linear dependence of the time necessary for STT switching versus the applied magnetic field. Both effects are entirely of dynamical nature and their influence on the phase boundaries can be theoretically described using the formalism developed in Ref. [16]. Renormalization of the effective dynamic time allows one to link dependence between the critical current, pulse width and finite temperature. This will be also done in the next section, while here the discussion will be focused on a qualitative analysis of the relative contributions of the in-plane and out-of-plane STT terms to phase boundaries shapes.

One can see from Fig.3a that the general behavior of the phase boundaries modification on the simulated phase diagrams under finite writing pulse regime is in agreement with the conclusions made in the previous sections for the DC regime. For the case of $s_{t\perp 2}$ = 400· $G_p^{-2}$ Oe/(Volt)$^2$ and $s_{t\parallel}$ = 0 · $G_p^{-1}$ Oe/Volt, the simulated phase diagram demonstrates a unidirectional STT switching due to quadratic dependence of $a_\perp$ versus applied voltage. In other words, only switching to the antiparallel configuration is possible for $s_{t\perp 2}$> 0, $s_{t\parallel}$ = 0. Zero-field ($H_{ext} = 0$) STT switching voltage for this diagram is =+/-1.6V. This voltage induces an effective STT field in the damping term of Eq.(2) ~1000 Oe, which is five times higher than the effective perpendicular anisotropy field $H_\perp$ =200 Oe. At the same time, if one adds a relatively small damping-like prefactor $s_{t\parallel}$ =30 · $G_p^{-1}$ Oe/Volt it completely removes any apparent influence of the field-like STT term from the phase diagram despite of the huge value chosen for its prefactor. When the effective contributions from both prefactors are comparable, the phase diagram acquires a noticeable asymmetry, as can be seen for the last two diagrams in the middle column. However, such combination of $s_{t\parallel}$ and $s_{t\perp 2}$ can be already physically unrealistic.

Figure 3b shows the same set of simulations made under T=300K conditions. Several temperature-induced effects are observed there: *i)* Decrease of the coercive field showing that thermally activated magnetization reversal takes place when the external magnetic field substantially lowers the effective barrier height in the system; *ii)* Shift of the voltage-driven parts of the boundaries towards lower switching voltages. Thermal fluctuations of the magnetic moment direction increases the probability to launch STT switching thanks to a thermally-induced misorientation between $\hat{m}$ and $\hat{p}$. This increases the initial STT amplitude and substantially decreases the switching time for a given writing pulse amplitude. This is consistent with earlier observations in STT-MRAM cells and with theoretical expectation of a

$I_c = I_{c0}\left\{1 - \frac{k_B T}{\Delta E} Ln\left(\frac{\tau}{\tau_0}\right)\right\}$ dependence of the switching current on the pulse duration under finite temperature [39]. Therefore, Fig.3b Indicates that the general features observed in the switching phase diagram at 0K (i.e. Fig.3a) are conserved at finite temperature and illustrates again the negligible role of the out-of-plane STT term in the switching process (see in particular the last column in Fig.3b).

### 5. Experimental measurements of the (I,H) switching diagram

In this section, the STT efficiency and other magnetic parameters of pMTJ pillars are directly extracted from the measured diagram. Nominal 50 nm diameter pMTJ pillars were fabricated from an MTJ stack grown by magnetron sputtering. The stack contains a 1.7nm thick $Co_{20}Fe_{60}B_{20}$ free layer sandwiched between two MgO barriers. Magnetization saturation parameter of the free layer was measured to be 1030 emu/cm$^3$. Current in-plane magnetotransport measurements (CIPTMR) yielded RxA = 5.7 Ω µm$^2$ and TMR=126 %. The second MgO barrier was introduced to increase the perpendicular anisotropy of the free layer. It has a negligible resistance-area (RA) product compared to the main tunnel barrier. The bottom fixed layer is a synthetic antiferromagnetic-based perpendicularly magnetized multilayer and the polarizer material has the same composition as for the free layer. The metallic electrode above the second MgO barrier is non-magnetic. Experimentally, it was found that the actual pillar diameter slightly differs from its nominal value due to the nanofabrication technology (36nm instead of 50nm nominal). This was recalculated using the values of the low resistive state ($R_{pp}$ = 5.6 kΩ) of the magnetoresistance curve (Fig.4a) and assuming that RxA value is preserved after the nanofabrication. Knowing the volume of the free layer in the pillar, $V_p$, its room temperature coercivity, measurement time (~1s) and attempt frequency $f_0$ = 10$^{10}$ s$^{-1}$, one can recalculate the perpendicular magnetic anisotropy from Neel-Brown formula [40,37]:

$$H_C(T) = H_\perp \left(1 - \sqrt{\frac{2\, k_B T \ln(t_m f_0)}{M_S H_\perp V_p}}\right), \qquad (10)$$

which gives $H_\perp$=2.6 kOe and Δ = 56.

The phase diagram measured at room temperature is shown in Fig. 4b. At each magnetic field point, a 100 ns writing pulse with fixed amplitude was applied to the pMTJ pillar. Subsequently, the resistance was measured under small DC bias current and the next magnetic field point was set. To reduce the stochasticity in the switching field values, the magnetoresistance loop was measured 15 times and their average was used for switching fields determination. The same procedure was used for all writing pulse amplitudes and the final phase diagram was constructed from these averaged magnetoresistance loops. Magnetic field loop repetition frequency was 2 Hz.

The extracted phase boundaries are shown in Fig.4c. The coercive field of the free layer is 940 Oe and the coupling field with the reference layer is only 11 Oe and it is ferromagnetic. The voltage driven parts are linear and almost parallel to each other. To reduce the influence of small nonlinearities at the edges of the boundaries, only the central parts (within +/- 500 Oe region) were used in the fitting. The extracted slopes are 1.27·10$^{-4}$ Volt/Oe and 1.23·10$^{-4}$ Volt/Oe, their difference is within the fitting error. The zero field switching voltages are 0.359 Volt and 0.385Volt respectively. The difference is most probably due to the small DC bias current used for the resistance measurements.

The phase diagram shape is similar to those obtained from the theoretical analysis (Sec.3) as well as from the simulations (Sec.4) where the out-of-plane STT term is not dominating. For this system, we can choose the STT prefactors model $a_\perp = 0$, $a_\parallel = s_{t\parallel} G_p V_{bias}$. It corresponds to DC diagram shown in Fig.2 whose boundaries are described by Eqs. (8,9). To recalculate $s_{t\parallel}$ parameter from the extracted diagram slopes, one needs firstly to remap the experimental finite temperature – finite writing pulse diagram to the model case of long pulse – low temperature diagram. Here, we will follow the formalism described in Ref.[16]. Thermal effects in our case can be reduced to the regime of thermally assisted ballistic STT switching. In this regime, the main role of thermal fluctuations is to increase the probability of STT switching thanks to increased initial misorientation angle $\theta_0$, $|\cos(\theta_0)| \neq 1$. As already mentioned, STT switching dynamics starting from a tilted state reduces the switching time $\tau$ in agreement with [13,14]. The cone angle, $2\theta_0$, for which the equilibrium probability for the magnetic moment orientation distribution is 0.5, is determined by thermal stability parameter $\Delta$ and applied magnetic field $\theta_0 = (\ln 2 / \Delta)^{1/2} (1 + H_{ext}/H_\perp)^{-1/2}$, while the final angle, the extremum on the energy barrier $\theta_\tau = \arccos(-H_{ext}/H_\perp)$ (for $\theta_0 < \pi/2$), is determined by magnetic field (see Eq.77 in Ref.[16]). Having defined the initial $\theta_0$ and final $\theta_\tau$ angles of the STT-induced dynamics, one can calculate analytically the switching time $\tau$ (see Eq. (58) in Ref[16]):

$$(i-1)\frac{\tau}{\tau_D} = \ln\left(\frac{x_\tau}{x_0}\right) - \frac{1}{i+1}\ln\left(\frac{\frac{i-1}{i+1}+x_\tau^2}{\frac{i-1}{i+1}+x_0^2}\right), \tag{11}$$

here $x_0 = \tan\theta_0$, $x_\tau = \tan\theta_\tau$, $\tau_D = \frac{(1+\alpha^2)}{\alpha\mu_0\gamma H_\perp}$ and, according to our formalism, $i = I_{sw}^\tau/I_{sw0} - \frac{H_{ext}}{H_\perp}$. Having calculated $\theta_0 = 6°$, $\tau_D = 9.9 \cdot 10^{-9}$ s and assuming $\alpha = 0.02$ [41] and writing pulse duration $\tau = 100 \cdot 10^{-9}$ s, we recalculated $I_{sw}^\tau(H_{ext})$ dependence from Eq.(11), which is blue line in Fig.5, and compared it with the case of the DC diagram $I_{sw}(H_{ext})$, which is shown by circles in Fig.5, derived from Eq.(8-9). One can conclude that 100ns writing pulses are long enough to remove the effect of dynamical distortion of the phase boundaries. For the measured device of Fig.5, we find $\frac{\tau}{\tau_D} = 100.6$ which is quite high. This gives the possibility to work directly with the phase boundaries (Eq.(8-9)) derived from Eq.(7). However, if $\frac{\tau}{\tau_D} < 10$ (if the writing pulse width in the experiment would be lower than 10 ns) and/or $\theta_0$ is too small, the phase boundaries remapping procedure is necessary before further analysis of the phase boundaries can be made. Indeed, in the simulations shown in the previous sections, the respective value of $\frac{\tau}{\tau_D}$ is 1.54. Therefore, the switching currents are much higher and the linear slope is different from that expected from the model. One also should notice that this formalism works only in high-Δ approximation. Therefore, the parts of the phase boundaries which are close to the regions where $H_{ext}$ approaches $H_\perp$ should be removed from the analysis.

From extrapolation of the voltage driven boundaries to V=0 one can estimate $H_\perp$ ~ 2.8-3.1 kOe, which is slightly higher than the corresponding value extracted from Eq.10 (2.6kOe). Nevertheless, the obtained $H_\perp$ values are in quite good agreement considering that these two values are derived from very different physical phenomena (superparamagnetism vs STT switching). The spin-torque efficiency prefactor, $s_{t\parallel}$, can be directly determined from the experimental slope using Eq.(9): $s_{t\parallel}$=162· $G_p^{-1}$ Oe/Volt. From this, assuming that $G_p$ =1/RxA, the effective spin polarization parameter in the system can be derived: $\eta$ = 0.49. If one uses the measured TMR value to estimate the polarization factor assuming that $\eta = \sqrt{TMR\,(TMR+2)}/(2(TMR+1))$ [42] and TMR = 1.26, this would yield $\eta$=0.44, which is close to the value extracted from the diagram boundary slope. The zero-field switching current, recalculated using Eq.(8) for obtained values of $H_\perp$, $s_{t\parallel}$ and known parameter $\alpha$ gives $I_{sw0}$=0.35 $G_p$ · Volt.

Therefore, one can conclude that the experiments carried out on 36 nm pMTJ system can be well described within the macrospin approximation and thermally activated ballistic regime of STT switching. $H_\perp$, $s_{t\parallel}$ parameters extracted from the phase boundaries of V$_{bias}$-H stability diagram are in good agreement with those extracted independently from magnetoresistance loop and Neel-Brown model.

## 6. Landau vs Gilbert

In this section, we emphasize an important issue naturally arising from the analysis carried out in the previous sections. If the STT terms are added directly into Landau-Lifshitz (LL) equation [43], then instead of Eq. 2 (obtained with Gilbert dissipation term [44]) the following modified equation is obtained:

$$\frac{1}{\gamma\mu_0}\frac{d\hat{m}}{dt} = -\hat{m} \times \left(\frac{1}{1+\alpha^2}\vec{H}_{eff} - a_\perp \hat{p}\right) - \hat{m} \times \left(\hat{m} \times \left(\frac{\alpha}{1+\alpha^2}\vec{H}_{eff} + a_\parallel \hat{p}\right)\right). \qquad (12)$$

Still preserving the main features and general behavior of the STT switching in fully perpendicular structures, Eq. 10 forbids the switching only by the out-of-plane STT-term, in contrast to Eq. 2 where the ($\alpha\,a_\perp \hat{m} \times (\hat{m} \times \hat{p})$) component allows the system to change its energy even if $a_\parallel = 0$. That turns us to the still open discussion [45-52] of physical validity of Gilbert-like damping and Landau-like damping formulation in the equation of magnetization dynamics. Although it is generally claimed that LL and LLG equations are mathematically equivalent, we can see a significant difference when the STT terms are added: field-like STT term written in LL equation is **_fully conservative_** and it cannot change the system energy if Eq.12 is chosen to describe the STT-induced dynamics. Leaving this fact "as is", one should notice that in numerical simulations, it is more common to use LL form instead of LLG form and different ways to introduce STT-terms (i.e. explicitly into LL equation (Eq.12) or via transformation of LLG+ STT (Eq.2)) can lead to significantly different results.

Figure 6 demonstrates this important issue by comparing examples of macrospin simulations using either Landau-Lifshitz or Gilbert damping terms to describe the dissipation during STT-induced switching. Here, we adjusted the relative magnitudes of the field-like and damping-like STT prefactors to have comparable contributions in the second part of Eq. 2, which is LLG+STT case. As soon as the field-like STT prefactor is set to have only a quadratic bias voltage dependence (the case of a symmetrical tunnel junction), the produced torque always pulls the free layer magnetization in the antiparallel configuration with the fixed layer. The damping-like STT prefactor is set to be linear on the bias voltage and therefore the torque direction is determined by the current polarity. When a negative voltage is applied to the system, field-like torque helps the damping-like torque to switch the magnetization in the antiparallel state. It shifts the phase boundary towards lower switching voltages. However the expected boundary shift is too small to be visible in our simulations considering the chosen step for the voltage writing pulse amplitude. Also a quadratic dependence of the field-like STT prefactor allows it to compete with the damping like torque only at relatively high writing pulse voltages. At the same time, for positive pulses, field-like torque works against the damping-like torque, which shifts the phase boundary to higher voltages. The higher the switching voltage – the higher the relative contribution from the field-like torque. Finally, when the writing pulse is about 1.6 V, field-like torque compensates the damping-like one and further increase of the writing pulse amplitude starts shifting the phase boundary back towards negative fields, decreasing the field window of the bipolar STT switching. The same effect is observed at finite temperatures on Fig. 3b for the bottom-middle diagram. This competition between the STT terms, however, is impossible in case of simulation with the Landau damping term because $\alpha\, a_\perp \hat{m} \times (\hat{m} \times \hat{p})$ term is absent in Eq.12.

Finally, it is traditionally accepted that Landau-Lifshitz-Gilbert and Landau-Lifshitz equations are geometrically equivalent and the mathematical transformation from one to another ends up with $\frac{1}{1+\alpha^2}$ rescaling of the gyromagnetic ratio. This $\frac{1}{1+\alpha^2}$ correction in real physical systems is very small and experimentally undetectable. However, this is not the case anymore if the STT terms are added to the LLG equation. The equations are now ***different.*** The same transformation (i.e. LLG+STT -> LL) leads to appearance of two additional STT pseudo-torques ($\alpha\, a_\perp \hat{m} \times (\hat{m} \times \hat{p})$, $\alpha\, a_\parallel \hat{m} \times \hat{p}$) which are linearly proportional to the damping constant $\alpha$ and in principle can be experimentally detected.

Experimentally, it should be possible to assess which formulation of damping is correct by measuring the variation of the precession frequency in the sub-switching threshold regime in samples having various damping constants. Such samples could be produced for instance by depositing a wedge of Pt above the storage layer before patterning of the wafer. For this experiment, it would be preferable to use symmetric MTJs so that the field like torque has a quadratic dependence on bias voltage. If the LLG formulation is correct, we expect a linear variation of the frequency with damping constant under fixed bias voltage whereas if the LL formulation is valid, no dependence of the frequency on damping should be observed.

## 7. Conclusions

It has been shown that Landau-Lifshitz-Gilbert equation with the field-like and damping-like STT terms transformed into the Landau-Lifshitz form considerably simplifies the analysis of the STT switching process. In case of a fully perpendicular MTJ system, the boundaries of the I-H stability phase diagram can be directly obtained from the transformed equation (2). It was shown that the field-like term has negligible influence on the STT switching process in pMTJs with low damping, influencing mainly the FMR precession frequency for the small oscillations near the equilibrium. Considering that in standard pMTJ structures its effective magnitude cannot be much higher than the magnitude of the in-plane torque, it would be hard to track its bias voltage (current) dependence from experimentally measured stability phase diagrams. Measuring the bias voltage dependence of the frequency in the precessional regime would certainly better reveal the influence of the field-like STT term but still the contribution of the field like term would have to be separated from the non-linear influence of the oscillations amplitude on the frequency.

Finite temperature macrospin simulations in LLG-STT formalism under finite writing pulse duration have confirmed the negligible role of the field-like term in the STT switching process of pMTJ structure. Limitations of the macrospin model are not expected to be important in the case of pMTJ pillars with diameter comparable to or below the exchange length. This is confirmed by the experiments which were carried out on 36 nm diameter pMTJ pillars.

One should note that the developed method for the phase boundaries construction gives the same results as those obtained from the analysis of dynamical response of the system, carried out by different groups supposing the linear dependence of the damping-like STT prefactor versus applied bias voltage. However, we believe that it will be more useful in the interpretation of the experiment and simulations, because it is much more flexible and it allows to introduce any desirable current (voltage) dependences for the in-plane and out-of-plane spin-torque prefactors.

Using the developed formalism, the spin-torque efficiency and effective spin polarization parameters have been derived from the current-field stability diagram boundaries experimentally measured on 36 nm pMTJ pillar. The obtained parameters have been cross-checked by estimations from magnetoresistance curves and from the thermally activated magnetization reversal regime. Good agreement between the values derived from the analysis of different physical principles strongly supports the assumption of macrospin-like behavior in the measured sample.

We also showed that the different dissipation terms (i.e. Landau-Lishfitz or Gilbert) give rise to different analytical expressions describing the phase boundaries of I-H switching diagrams, which can be important in heavily damped systems. If Landau damping term is physically correct, the action of the field-like and the damping-like torques in pMTJ system is completely separated in precession and dissipation terms in the equation of dynamics. While if Gilbert damping term is correct, then two additional torques ($\alpha\, a_\perp \hat{\boldsymbol{m}} \times (\hat{\boldsymbol{m}} \times \hat{\boldsymbol{p}})$ and

$\alpha\, a_\parallel\, \hat{\boldsymbol{m}} \times \hat{\boldsymbol{p}}$) are mixed up to the main STT contributors ($a_\parallel\, \hat{\boldsymbol{m}} \times \hat{\boldsymbol{p}}$ and $a_\perp\, \hat{\boldsymbol{m}} \times (\hat{\boldsymbol{m}} \times \hat{\boldsymbol{p}})$ respectively). An experimental way to assess which damping formulation is correct in combination with STT was proposed.

**Acknowledgements**

This work was supported by the Samsung Global MRAM Innovation Program and EUROTALENTS Program. The authors are also grateful to Ursula Ebels for fruitful discussions.

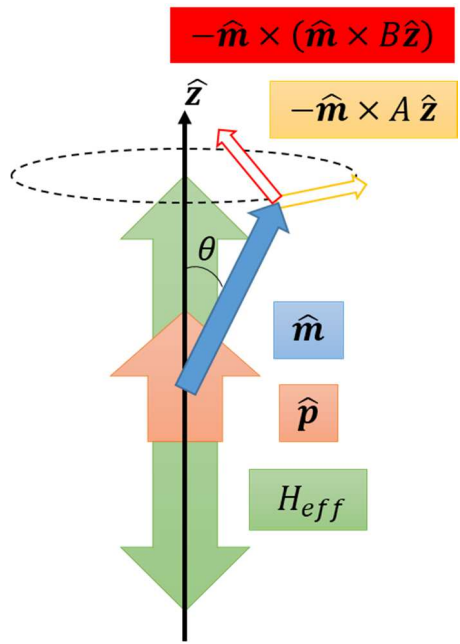

Fig.1 Geometry of the fully perpendicular MTJ system.

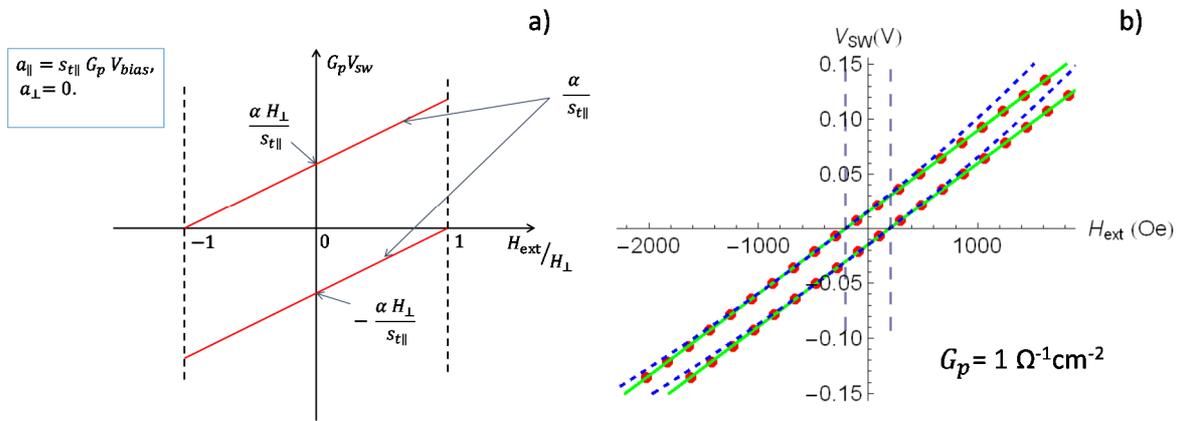

Fig.2 (a) - Stability phase diagram constructed from Eq.(7) assuming $a_\parallel = s_{t\parallel} G_p V_{bias}$ and $a_\perp = 0$; (b) – Modification of the phase boundaries for the same $a_\parallel$ prefactor ($a_\parallel = s_{t\parallel} G_p V_{bias}$, $s_{t\parallel}$ = 67 · $G_p^{-1}$ Oe/Volt) and different forms of $a_\perp$ prefactor: solid line $a_\perp = 0$ ; circles $a_\perp = s_{t\perp 2} (G_p V_{bias})^2$ with $s_{t\perp 2}$ = 154 · $G_p^{-2}$ Oe/(Volt)$^2$; dashed line $a_\perp = s_{t\perp 1} G_p V_{bias} + s_{t\perp 2} (G_p V_{bias})^2$ with $s_{t\perp 1}$ = 500 · $G_p^{-1}$ Oe/Volt and $s_{t\perp 2}$ = 10000 · $G_p^{-2}$ Oe/(Volt)$^2$; Other system parameters are: $\alpha$ = 0.05, $H_\perp$ = 200 Oe.

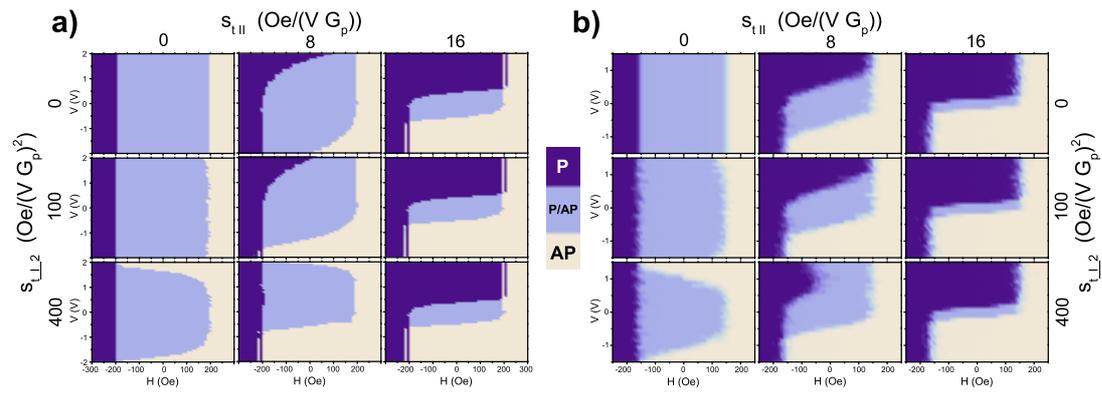

Fig. 3. Finite writing pulse phase diagrams for different in-plane and out-of-plane STT prefactors magnitudes. The model parameters $H_\perp$ = 200 Oe, g =2.20 (g-factor), $\alpha$ = 0.01. Integration time was 1 microsecond in each field point and the writing pulse width is 40 ns. a) T = 0K case, the axes scale is the same for all diagrams: +/- 2 V from top to bottom and +/- 300 Oe from right to left. b) T=300K, the axes scale is: +/- 1.5 V and +/- 250 Oe.

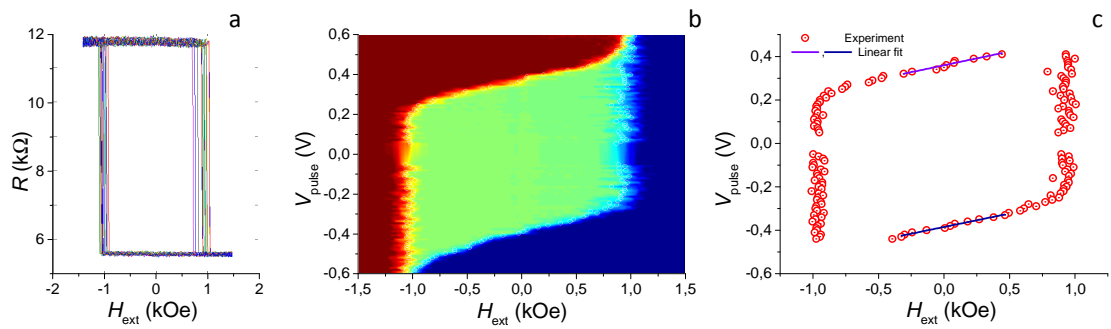

Fig. 4. Experiment carried out on pMTJ pillar at room temperature applying 100 ns writing pulses. a) Examples of magnetoresistance loops measured with zero writing pulses; b) Stability phase diagram; c) Extracted phase boundaries and their linear fittings.

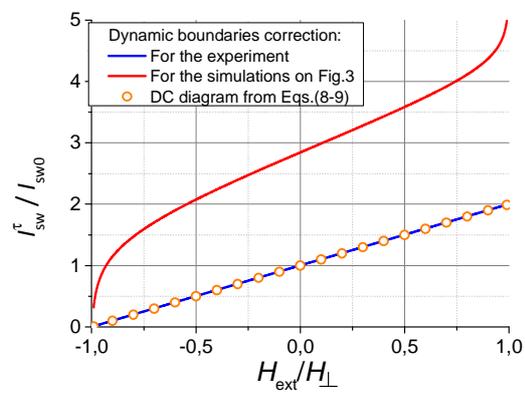

Fig. 5. Finite pulse – finite temperature diagram boundary for $\frac{\tau}{\tau_D}$ =100.6 (for experiment - blue) and $\frac{\tau}{\tau_D}$ =1.5 (for simulations - red). The dots are respective boundary obtained from Eqs. (8-9).

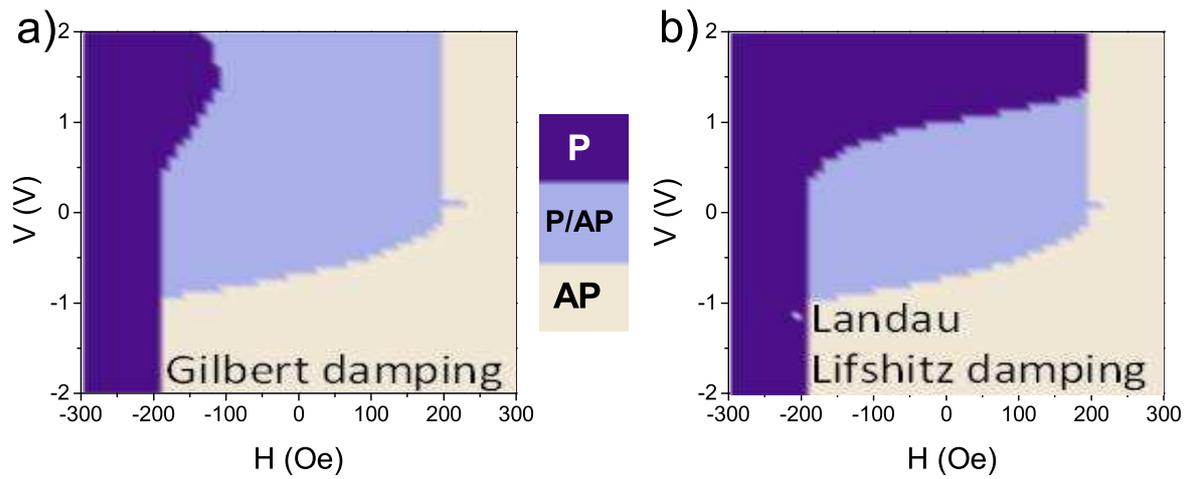

Fig. 6. Two identical macrospin simulations of a stability phase diagram carried at T = 0K: (a) using Eq.12 (LLG + STT) ; (b) using Eq.2 (LL + STT). STT prefactors: $s_{t\parallel} = 12 \cdot G_p^{-1}$ Oe/Volt and $s_{t\perp 2} = 400 \cdot G_p^{-2}$ Oe/(Volt)$^2$. Other parameters are the same as used for the simulations in Section 4.